\documentclass{article}%
\usepackage{amsmath}
\usepackage{authblk}
\usepackage{amsfonts}
\usepackage{amssymb}
\usepackage{graphicx}%
\usepackage{lineno}

\providecommand{\U}[1]{\protect\rule{.1in}{.1in}}

\begin{document}

\title{Violation of contextual generalization of the Leggett-Garg inequality for recognition of ambiguous figures}

\author[1]{Masanari Asano}
\author[2]{Takahisa Hashimoto}
\author[3]{Andrei Khrennikov}
\author[2]{Masanori Ohya}
\author[2]{Yoshiharu Tanaka}

\affil[1]{Liberal Arts Division, Tokuyama College of Technology,\\
 Gakuendai, Shunan, Yamaguchi 745-8585 Japan }

\affil[2]{ Department of Information
Sciences, Tokyo University of Science, \\
 Yamasaki 2641, Noda-shi, Chiba, 278-8510 Japan}

\affil[3]{International Center for Mathematical Modeling \\
in Physics, Engineering, Economics, and Cognitive Science\\
Linnaeus University, V\"axj\"o-Kalmar, Sweden}

\date{}
\maketitle

\abstract{We interpret the Leggett-Garg (LG) inequality as a kind of contextual probabilistic inequality in which 
one combines data collected in experiments performed for three different contexts. In the original version of the 
inequality these contexts have the temporal nature and they are given by three pairs of instances of time,
$(t_1, t_2), (t_2, t_3), (t_3, t_4),$ where $t_1 < t_2 < t_3.$  We generalize LG conditions of macroscopic realism and 
noninvasive measurability in the general contextual framework. Our formulation is done in the purely probabilistic terms:  existence 
of the context independent joint probability distribution $P$ and the possibility to reconstruct the experimentally found marginal (two dimensional)
probability distributions from the $P.$ We derive an analog of the LG inequality, ``contextual LG inequality'', and use it 
as a test of ``quantum-likeness'' of statistical data collected in a series of experiments on recognition of   ambiguous figures. 
In our experimental study the figure under recognition is the Schr\"oder stair which is shown with rotations for different angles.  
Contexts are encoded by dynamics of rotations: clockwise, anticlockwise, and random. Our data demonstrated violation of the 
contextual LG inequality for some combinations of aforementioned contexts. Since in quantum theory and experiments with quantum physical
systems this inequality is violated, e.g., in the form of the original LG-inequality, our result can be interpreted as a sign that 
the quantum(-like) models can provide a more adequate description of the data generated in the process of recognition of ambiguous figures.}

\section{Introduction}

Mathematical modeling of the process of recognition of ambiguous figures is an intriguing problem which still has no completely satisfactory 
solution. Recently quantum(-like) models based on the mathematical formalism of quantum mechanics and its generalizations
were applied to this problem \cite{C0}--\cite{ATM1}. As in any mathematical modeling project, the output of models has to be compared with results of 
experiments. 

One of the basic intrinsically quantum probabilistic effects is {\it violation of the formula of total probability} which is experimentally 
exhibited in the interference effect \cite{K1}--\cite{NEW}. In a series of papers \cite{C0}--\cite{C2} such an effect was found 
in experimental data collected in sequential recognition 
of a pair of ambiguous figures. Then it was also found  \cite{C3} that these data violate Bell's inequality \cite{B}. 

Another quantum-like study on recognition of ambiguous figures was done by Atmanspacher {\it et al.}
\cite{ATM}, \cite{ATM0} a quantum-like model of bistable perception.  (A generalized quantum formalism was in use \cite{ATM1}.) 
One of the important novelties in \cite{ATM}, \cite{ATM0} was application of {\it temporal Bell inequalities}, 
concretely the Leggett-Garg (LG) inequality \cite{LG}. 

Recently a quantum-like model of recognition 
of ambiguous figures was presented in the paper of Asano {\it et al.} \cite{AS}.  The model matches very well with experimental data    
on determination of the structure of the Schr\"oder stair which was shown with rotations for different angles. 

In this paper  we consider the possibility to use the inequalities of the LG type to check ``quantum-likeness'' of statistical data
collected \cite{AS} in experimental studies on bistable perception.
We interpret the LG inequality as a kind of contextual probabilistic inequality in which 
one combines data collected in experiments performed for three different contexts, cf. \cite{KHR_CONT}, \cite{K1}, \cite{ATM}, \cite{Dzhafarov}, 
\cite{NEW}.  In the original version of the 
inequality these contexts have the temporal nature and they are given by three pairs of instances of time,
$(t_1, t_2), (t_2, t_3), (t_3, t_4),$ where $t_1 < t_2 < t_3.$  We generalize LG conditions of macroscopic realism and 
noninvasive measurability in the general contextual framework. Our formulation is done in purely probabilistic terms: existence 
of the context independent joint probability distribution $P$ and the possibility to reconstruct the experimentally found marginal (two dimensional)
probability distributions from the $P.$ We derive an analog of the LG inequality, ``{\it contextual LG inequality}'', and use it 
as a test of ``quantum-likeness'' of statistical data collected in a series of experiments on recognition of   ambiguous figures. 
In our experimental study the figure under recognition is the {\it Schr\"oder stair}  \cite{b0} which is shown with rotations for different angles.  
Contexts are encoded by dynamics of rotations: clockwise, anticlockwise, and random. Our data demonstrated violation of the 
contextual LG inequality for some combinations of aforementioned contexts. Since in quantum theory and experiments with quantum physical
systems this inequality is violated, e.g., in the form of the original LG-inequality (see, e.g., \cite{Vi} and references hereby), 
our result can be interpreted as a sign that 
the quantum(-like) models can provide a more adequate description of the data generated in the process of 
recognition of ambiguous figures.

In probabilistic terms context dependence of (probabilistically determined) mental states imply that the conventional model of probability 
theory, the Kolmogorov measure-theoretic model \cite{K}, cannot be applied to describe statistics of recognition of ambiguous figures. Thus 
our result on violation of the contextual LG inequality restricts the domain of applications of the Kolmogorov model\footnote{In fact, the real 
situation is more complicated, see section \ref{HHH}.}. More general
probabilistic models have to be tested, e.g., quantum probability and its generalizations \cite{ATM1}, \cite{K1}, \cite{OH}. Since neither physicists nor 
psychologists have a proper education in probability theory, at least in the axiomatic approach, we complete this paper with an extended
appendix presenting both a brief introduction to the classical Kolmogorov model and a discussion on possible non-Kolmogorovian generalizations
as well as relation to Bell type inequalities.     

Contextuality of mental representations is one of the fundamental features of cognition. In particular, mental contextuality is one of the main
motivations for applications of the quantum formalism to modeling of cognition  \cite{K1},\cite{OH}--\cite{OH2}, \cite{AS}
and more generally biological information 
processing\footnote{The formal identity of quantum-like models of ``decision making'' by cells and cognitive systems can be considered as 
an argument for  
recognition of existing of a kind of cell's cognition.} \cite{OH3}, \cite{OH4},    since quantum mechanics is also 
fundamentally contextual. Typically quantum contextuality is expressed in the form of the Kochen-Specker theorem. However, recently contextuality 
was also represented with the aid of Bell type inequalities, e.g., \cite{Rauch}. These recent theoretical and experimental results on the Bell 
type expression of contextuality match well with the contextual approach to the problem of violation of Bell's inequalities developed by one 
of the authors of this paper \cite{KHR_CONT}, \cite{KHR_CONT1}--\cite{KHR_CONT2}, see also \cite{Accardi}--\cite{DZ2}.
Originally Bell mixed in one cocktail nonlocality and realism. 
The standard conclusion from violation of  the inequalities
of the CHSH-type is that ``local realism'' is incompatible with quantum behavior is not easy to interpret. What is the problem? Nonlocality? Realism? Both? 
The contextual viewpoint to violation of Bell type (so to say, spatial) inequalities helps a lot to clarify this problem.   
The contextual (spatial) Bell inequality is violated for a single particle, e.g., a neutron \cite{Rauch}, therefore the problem of (non)locality
can be automatically excluded from consideration. The same happens in the case of temporal contextual Bell inequalities, including the LG inequality.
We shall come back to this discussion and extend it to the problem of inter-relation of mental contextuality and mental realism in sections
\ref{COND},  \ref{concl}. (We also remark that T.  Nieuwenhuizen invented the terminology {\it Contextuality Loophole} summarizing outputs 
of studies  \cite{Accardi},  \cite{KHR_CONT}, \cite{KHR_CONT1}--\cite{KHR_CONT2}, \cite{Marian}-\cite{Raedt3}.) 
  
We point out that contextual dependent systems (in physics, biology, economics, finances, and social science) and non-Kolmogorov probability theory
can be described mathematically \cite{OH}, \cite{NEW}  by the adaptive dynamics and the operation of lifting (the latter is widely used in quantum 
information theory \cite{VOH}). 
  
It is important to remark that we consider contextuality in its the most general form, as
N. Bohr \cite{BR} did: the whole experimental arrangement has to be taken into account. J. Bell \cite{B} considered only a very special form of contextuality:
dependence of the result of measurement of some observable $A$  on joint measurement of another observable $B$ compatible with $A.$ In cognitive science
the situation is even trickier. Measurements are typically {\it self-observations} which the brain performs on itself. Therefore ``the whole 
experimental arrangement'' includes not only the ``external experimental arrangemenet'', e.g., prepared by researchers in cognitive psychology, 
but also ``internal arrangement'' including the brain state. 
  
We remark that this paper has nothing to do with study of quantum physical processes in the brain. We proceed with the operational 
approach to quantum theory as a formalism describing in probabilistic terms measurements of in general incompatible observables. Such 
observables can be of any nature, physical, mental, biological \cite{K1}, \cite{OH}.      

\section{Leggett-Garg inequality}

\subsection{Conditions of derivation}
\label{COND}

At the beginning of the discussion in the paper \cite{LG}, 
Leggett and Garg (LG) postulated the following two assumptions:

\begin{enumerate}
\item[(A1)] \textbf{Macroscopic realism}\textit{: A macroscopic system with
two or more macroscopically distinct states available to it will at all times
be in one or the other of these states.}

\item[(A2)] \textbf{Noninvasive measurability}\textit{: It is possible, in
principle, to determine the state of the system without arbitrarily small
perturbation on its subsequent dynamics. }
\end{enumerate}

Under these assumptions, the correlation functions must satisfy the 
LG inequality which will be presented in the next section. However, quantum mechanics
violates the LG inequality as well as the same analogue of Bell's inequality or
CHSH inequality. Therefore this violation means that at least one of the two
assumptions fails for quantum systems. 

Although in the derivation of the LG inequality, section \ref{DER}, both conditions 
play important roles, their foundational value is different. The main issue is realism,
whether one can still proceed with (A1), macroscopic realism, in the quantum world. 
Therefore the main part of the LG paper \cite{LG} is devoted to discussion about possible 
physical experimental schemes which may lead to noninvasive measurements or at least 
measurements in which invasiveness is small comparing with the degree of violation of 
the LG inequality. There are claims,
e.g., the experiment in \cite{Vi}, that such negligibly invasive measurements were performed 
experimentally and the LG inequality was violated. This is often seen as an important argument in favor of non-objectivity
of quantum observables. 

However, the LG approach plays an important foundational role even if the possibility that measurements are non-negligibly 
invasive cannot be excluded. We know that classical systems and measurements on such systems satisfy conditions (A1) and (A2);
e.g., airplane's trajectory. Therefore by violating LG we at least know that a phenomenon under study cannot be described 
classically.

In cognitive science it is not easy (if possible at all) to come with an experimental scheme which would lead to (at least 
approximately) noninvasive measurements. The brain is a kind of self-measurement device, by giving an answer to a question 
the brain definitely perturbs its mental state and non-negligibly. And the introspective measurements have definitely the lowest 
degree of non-invasiveness. Therefore it seems that violations of  the LG type inequalities for data collected, e.g., in cognitive 
psychology, cannot lead to the conclusion that mental realism is questionable. Here realism is understood in the sense of objectivity 
of mental observables, that their values can be assigned, say to the brain, a priori, i.e., before measurements.  Nevertheless, such violations
show that the data under consideration is nonclassical, i.e., it is not similar to data collected, e.g., from an ensemble of moving airplanes.       

However, our main point is that in relation to the problem of cognition the standard physical viewpoint on conditions for derivation 
of the LG inequality, namely, the mixture of macrorealism and non-invasiveness, does not match so well with the mental situation. As was emphasized,
the very notion of (non)invasive measurement loses its clearness for self-measuring devices and the brain is one of such devices. The paper
advertises the contextual viewpoint on the mental phenomena developed in the series of works  \cite{K1},\cite{OH}--\cite{OH2}, \cite{AS}. 
It seems that 
Bell type inequalities, including the temporal ones, can be used to distinguish 
contextual and non-contextual realism and more generally (since mental processes are fundamentally random) contextual and non-contextual 
probabilistic representations. As will be seen from coming presentation, non-contextuality of representation of probabilistic data implies
constraints on such data, in the form of various inequalities. By using contextual representation a system (including the brain) can violate 
such constraints. We shall come back to this discussion in section \ref{concl}. 
    
\section{Contextual viewpoint on the proof of LG inequality}
\label{DER}

To provide to a reader the possibility to compare the original LG inequality with our contextual generalization, see
section \ref{CGEN}, and at the same time to 
add the contextual flavor to the LG approach, we present the original LG derivation by considering time as a context parameter.  

Let $Q$ be an observable quantity which takes either $+1$ or $-1$. In
the original discussion by LG, $Q$ is the observable of position of a particle
in the two potential wells. However we can discuss another two-level system,
e.g. spin-$\frac{1}{2}$ system.

The measurement of the two-level system\ is perfomed on a single system at
different times $t_{1}<t_{2}<t_{3}$. We denote the observable at time $t_{k}$
by $Q_{k}\ (k=1,2,3)$.\ By repeating a series of three measurements, we can
estimate the value of correlation functions by
\[
C_{ij}=\frac{1}{N}\sum_{n=1}^{N}q_{i}^{(n)}q_{j}^{(n)},
\]
where $q_{i}^{(n)}$ (or $q_{j}^{(n)}$) is a result of the $n$-th measurement
of $Q_{i}$ (or $Q_{j}$). Note that the correlation between $Q_{i}$ and $Q_{j}$
takes the maximum value $C_{ij}=1$ when $q_{i}^{(n)}q_{j}^{(n)}\ $equals to $1$
for all the repeated trials. Here, consider the assumption A1, then the state
of the system is determined at all times even when the measurement does not
perform on the system. Therefore, the values of joint probabilities of
$Q_{1},Q_{2}$ and $Q_{3}$ are determined \textit{a priori} at initial time
$t_{0}$. We denote it by the symbol $P_{i,j}\left(  Q_{1},Q_{2},Q_{3}\right)  $. Remark
that the pairs of indexes $i,j$ encode the situation that only two observables $Q_{i}$ and $Q_{j}$
are measured. In other words, the joint probability depends on the situations
which pairs of observables  are measured. (We can consider pairs of indexes, instances of time, 
as parameters encoding three temporal contexts, ${\cal C}_{t_1t_2}, {\cal C}_{t_1t_3}, {\cal C}_{t_2t_3},$ 
cf. section \ref{CGEN}.) However if one considers (A2), then the
joint probabilities do not depend on temporal contexts\footnote{We remark that conditions (A1) and (A2) were formulated
in the physical framework. Therefore any study about the LG-inequality performed at the mathematical level 
of rigorousness has to present some mathematical formalization of these conditions. In our study (A1) and (A2) imply that there exists 
the joint probability distribution which does not depend on experimental contexts. In particular, we identify (A2) 
with {\it noncontextuality}. We understand well that this is not the only possible probabilistic 
interpretation  of (A1) and (A2).}:
\[
P_{i,j}\left(  Q_{1},Q_{2},Q_{3}\right)  =P\left(  Q_{1},Q_{2},Q_{3}\right)
\ \ \ \forall i,j
\]
Then we have the following equalities:
\begin{align*}
P\left(  Q_{1},Q_{2}\right)   &  =\sum_{Q_{3}=\pm1}P\left(  Q_{1},Q_{2}%
,Q_{3}\right)  ,\\
P\left(  Q_{2},Q_{3}\right)   &  =\sum_{Q_{1}=\pm1}P\left(  Q_{1},Q_{2}%
,Q_{3}\right)  ,\\
P\left(  Q_{1},Q_{3}\right)   &  =\sum_{Q_{2}=\pm1}P\left(  Q_{1},Q_{2}%
,Q_{3}\right) 
\end{align*}
which are consequences of the additivity of classical (Kolmogorov) probability. Thus pairwise joint probability 
distributions are context independent (as a consequence of  (A2)). We also have 
\begin{equation}
\label{e7}
P\left(  Q_{1}\right)   = \sum_{Q_{2}=\pm1} P\left( Q_{1},Q_{2}\right) = \sum_{Q_{3}=\pm1} P\left( Q_{1},Q_{3}\right);
\end{equation}
\begin{equation}
\label{e71}
P\left(  Q_{2}\right)   = \sum_{Q_{1}=\pm1} P\left( Q_{1},Q_{2}\right)=  \sum_{Q_{3}=\pm1} P\left( Q_{2},Q_{3}\right);
\end{equation}
\begin{equation}
\label{e72}
P\left(  Q_{3}\right)   = \sum_{Q_{1}=\pm1} P\left( Q_{1},Q_{3}\right) = \sum_{Q_{2}=\pm1} P\left( Q_{2},Q_{3}\right);
\end{equation}
Thus, for each observable, its probability distribution  is also context independent (as a consequence of (A2)).
Violation of these equalities is interpreted as exhibition of contextuality.  
In psychology and cognitive science the equalities (\ref{e7})--(\ref{e72}) represent the special case of 
so-called  {\it marginal selectivity} \cite{DZ}--\cite{DZ2}. It is clear that if at least one of these equalities is violated then one cannot assume existence of context independent joint probability 
distribution.

\medskip

Under the assumption of existence of the joint (triple) probability distribution 
the correlation functions are written with the joint probabilities
$P\left(  Q_{i},Q_{j}\right)  $\ as
\begin{align*}
C_{ij}  &  =P\left(  Q_{i}=1,Q_{j}=1\right)  +P\left(  Q_{i}=-1,Q_{j}
=-1\right)  \\
&  -P\left(  Q_{i}=-1,Q_{j}=1\right)  -P\left(  Q_{i}=1,Q_{j}%
=-1\right) \\
&  =2\left\{  P\left(  Q_{i}=1,Q_{j}=1\right)  +P\left(  Q_{i}=-1,Q_{j}%
=-1\right)  \right\}  -1.
\end{align*}
We set $K= C_{12}+C_{23}-C_{13}.$ It can be represented in the following form:
\begin{equation}
\label{LBLB}
K= 1-4\left\{  P\left(  Q_{1}=1,Q_{2}=-1,Q_{3}%
=1\right)  +P\left(  Q_{1}=-1,Q_{2}=1,Q_{3}=-1\right)  \right\}
\end{equation}
This representation implies the LG-inequality:
\begin{equation}
\label{LBLB1}
K \leq 1.
\end{equation}
As we know, e.g., \cite{LG}, \cite{Vi} for the quantum correlation functions $C_{ij}$ the above
inequality  can be violated (theoretically and experimentally)

\section{Contextual LG inequality}

\label{CGEN}

Here, we express the LG's assumptions in terms of  context-dependent probabilities \cite{KHR_CONT}.
We remark that in  general context-dependent probabilities cannot be represented in common Kolmogorov probability 
space. Therefore one can consider such contextual probabilistic models as non-Komogorovian probabilistic models, see the
appendix.    

\begin{enumerate}
\item[(A1)] There exists a joint probability $P_{\mathcal{C}}\left(
Q_{1},Q_{2},Q_{3}\right)  $ under a certain conditon of experiments (context)
$\mathcal{C}$. And the Kolmogorovness of $P_{\mathcal{C}}\left(  Q_{1}%
,Q_{2},Q_{3}\right)  $ is ensured within the context $\mathcal{C}$:
\begin{align*}
P_{\mathcal{C}}\left(  Q_{1},Q_{2}\right)   &  =\sum_{Q_{3}=\pm1}%
P_{\mathcal{C}}\left(  Q_{1},Q_{2},Q_{3}\right)  ,\\
P_{\mathcal{C}}\left(  Q_{2},Q_{3}\right)   &  =\sum_{Q_{1}=\pm1}%
P_{\mathcal{C}}\left(  Q_{1},Q_{2},Q_{3}\right)  ,\\
P_{\mathcal{C}}\left(  Q_{1},Q_{3}\right)   &  =\sum_{Q_{2}=\pm1}%
P_{\mathcal{C}}\left(  Q_{1},Q_{2},Q_{3}\right)  ,
\end{align*}
and%
\begin{align*}
P_{\mathcal{C}}\left(  Q_{1}\right)   &  =\sum_{Q_{2}=\pm1}P_{\mathcal{C}%
}\left(  Q_{1},Q_{2}\right)  =\sum_{Q_{3}=\pm1}P_{\mathcal{C}}\left(
Q_{1},Q_{3}\right)  =\sum_{Q_{2}=\pm1}\sum_{Q_{3}=\pm1}P_{\mathcal{C}}\left(
Q_{1},Q_{2},Q_{3}\right)  ,\\
P_{\mathcal{C}}\left(  Q_{2}\right)   &  =\sum_{Q_{1}=\pm1}P_{\mathcal{C}%
}\left(  Q_{1},Q_{2}\right)  =\sum_{Q_{3}=\pm1}P_{\mathcal{C}}\left(
Q_{2},Q_{3}\right)  =\sum_{Q_{1}=\pm1}\sum_{Q_{3}=\pm1}P_{\mathcal{C}}\left(
Q_{1},Q_{2},Q_{3}\right)  ,\\
P_{\mathcal{C}}\left(  Q_{3}\right)   &  =\sum_{Q_{2}=\pm1}P_{\mathcal{C}%
}\left(  Q_{2},Q_{3}\right)  =\sum_{Q_{1}=\pm1}P_{\mathcal{C}}\left(
Q_{1},Q_{3}\right)  =\sum_{Q_{1}=\pm1}\sum_{Q_{2}=\pm1}P_{\mathcal{C}}\left(
Q_{1},Q_{2},Q_{3}\right)  .
\end{align*}

\item[(A2)] Consider three different contexts $\mathcal{C}_{A},\mathcal{C}%
_{B}$ and $\mathcal{C}_{C}$, then there exists a context $\mathcal{C}$
unifying the above contexts $\mathcal{C}_{A},\mathcal{C}_{B}$ and
$\mathcal{C}_{C}$ such that
\begin{align*}
P_{\mathcal{C}_{A}}\left(  Q_{1},Q_{2}\right)   &  =\sum_{Q_{3}=\pm
1}P_{\mathcal{C}}\left(  Q_{1},Q_{2},Q_{3}\right)  ,\\
P_{\mathcal{C}_{B}}\left(  Q_{2},Q_{3}\right)   &  =\sum_{Q_{1}=\pm
1}P_{\mathcal{C}}\left(  Q_{1},Q_{2},Q_{3}\right)  ,\\
P_{\mathcal{C}_{C}}\left(  Q_{1},Q_{3}\right)   &  =\sum_{Q_{2}=\pm
1}P_{\mathcal{C}}\left(  Q_{1},Q_{2},Q_{3}\right)  .
\end{align*}
\bigskip
\end{enumerate}

From these assumptions, one can obtain the inequality (\ref{LBLB})  for $K$ given by
\begin{equation}
\label{Inequality}
K=   1-4( P_{\mathcal{C}}\left(  Q_{1}=1,Q_{2}=-1,Q_{3}=1\right)  +
P_{\mathcal{C}}\left(  Q_{1}=-1,Q_{2}%
=1,Q_{3}=-1\right)).  
\end{equation}

\section{Violation of  inequality in optical illusions}

\begin{figure}[htpb]
\centering{}\includegraphics[scale=0.3]{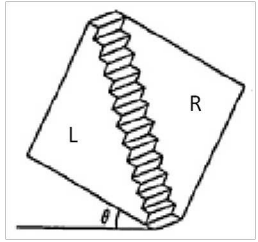} \caption{Schr\"oder's stair leaning at angle $\theta$}

\label{fig1} 
\end{figure}

The Schr\"oder's stair is an ambiguous figure which induces optical
illusion\cite{b0,b15}, see the Fig.~\ref{fig1}. Our brain can switch
between the two alternative interpretations of this figure:

\medskip

 (i) The surface of `L' is front, and the surface of `R' is back.

 (ii) The surface of `R' is front, and the surface of `L' is back. 

\medskip

This switch-like process of human perception is called \textit{depth inversion},
and many experimental proofs on this phenomenon have been reported.
However, the details of its mechanism is not completely figured out
even in recent studies. 

It is well-known fact that the depth inversion depends on various
contexts of figure; e.g. relative size of the surface `L' for `R',
color or shadow in figure, angle to the horizon, etc.\cite{b15}.
Therefore we must define the contextual dependent probability that
a person answers either (i) or (ii) in the experiment.

Now we explain the method of our experiment, and we show
its results. We show the subjects the picture of Schr\"oder's stair
which is leaning at a certain angle $\theta$ (see Fig.~\ref{fig1}).
We prepare the 11 pictures which are leaning at different angles:
$\theta=0,10,20,30,40,45,50,60,70,80,90$. A subject must answer either
(i)``L is front\textquotedblright{}\ or (ii)``R is front\textquotedblright{}\ for
every picture. We arrange the computer experiment to change the pictures
and to record their answers.

Before the experiment, we divided the subjects into three groups:
(A) 55 persons, (B) 48 persons, (C)\ 48 persons.\footnote{We randomly selected the 151 participants from the 
students of Tokyo University of Science. And randomly divided this sample into three aforementioned groups.
We showed a suite of 11 pictures to each subject with a laptop computer. Usually the 
distance between eyes and screen is 20 cm ~ 30 cm. He or she typed a key corresponding to answer $L$ or $R.$ 
We did not limited time to answer, so that the subject had enough time to make his or her decision.
We implemented java application in order to show the pictures and to record their answers. 
The order of showing in random case can be selected by software (not manually). 
All the records are saved as the output CSV file.} For the first group
(A), the order of showing is randomly selected for each person. To
assume statistically uniform randomness of this selection, we use
the computer-implemented function (e.g. java.rand). For the second
group (B), the angle $\theta$  is increased from a small  value:
$0,10,\cdots,90$. Inversely, for the third group (C), the angle $\theta$
is decreased from a large value: $90,80,\cdots,0$.

Thus we have the three kinds of experimental data: (A) the angle of Schr\"oder stair
changes randomly, (B) from 0 to 90, (C) from 90 to 0. These contexs of
experiments are denoted by $\mathcal{C}_{A},\mathcal{C}_{B}$ and
$\mathcal{C}_{C}$. Let $X_{\theta}$ be a random variable which takes $\pm1$.
The event that a subject says "left side is front" corresponds to the result
$X_{\theta}=+1$. Then, from the repeated trials for each experimental context,
we have the experimentally obtained values of joint probabilities:%
\[
P_{\mathcal{C}_{A}}\left(  X_{0},X_{10},\cdots,X_{90}\right)
,\ \ \ P_{\mathcal{C}_{B}}\left(  X_{0},X_{10},\cdots,X_{90}\right)
,\ \ \ P_{\mathcal{C}_{C}}\left(  X_{0},X_{10},\cdots,X_{90}\right)  .
\]
The correlation functions is given by%
\begin{align*}
C_{12} &  =2\left\{  P_{\mathcal{X}}\left(  X_{\theta_{1}}=1,X_{\theta_{2}%
}=1\right)  +P_{\mathcal{X}}\left(  X_{\theta_{1}}=-1,X_{\theta_{2}%
}=-1\right)  \right\}  -1,\\
C_{23} &  =2\left\{  P_{\mathcal{Y}}\left(  X_{\theta_{2}}=1,X_{\theta_{3}%
}=1\right)  +P_{\mathcal{Y}}\left(  X_{\theta_{2}}=-1,X_{\theta_{3}%
}=-1\right)  \right\}  -1,\\
C_{13} &  =2\left\{  P_{\mathcal{Z}}\left(  X_{\theta_{1}}=1,X_{\theta_{3}%
}=1\right)  +P_{\mathcal{Z}}\left(  X_{\theta_{1}}=-1,X_{\theta_{3}%
}=-1\right)  \right\}  -1.
\end{align*}
Here, the triple $\left(  \mathcal{X},\mathcal{Y},\mathcal{Z}\right)  $ is
given by a combination of the contexts $\mathcal{C}_{A},\mathcal{C}_{B}$ and
$\mathcal{C}_{C}$. We show the values of $C_{12},C_{23}$ and $C_{13}$ in the
Table XX\ 

\bigskip

\bigskip

\begin{center}%
\begin{tabular}
[c]{cccc}
& $C_{12}$ & $C_{23}$ & $C_{13}$\\
$\mathcal{C}_{A}$ & $1.000$ & $0.964$ & $0.964$\\
$\mathcal{C}_{B}$ & $0.917$ & $0.833$ & $0.750$\\
$\mathcal{C}_{C}$ & $0.917$ & $1.000$ & $1.000$%
\end{tabular}
\ \
\begin{tabular}
[c]{cccc}
& $C_{12}$ & $C_{23}$ & $C_{13}$\\
$\mathcal{C}_{A}$ & $0.091$ & $0.091$ & $0.127$\\
$\mathcal{C}_{B}$ & $0.375$ & $0.625$ & $0.083$\\
$\mathcal{C}_{C}$ & $0.625$ & $0.375$ & $0.167$%
\end{tabular}

(Left) $\left(  \theta_{1},\theta_{2},\theta_{3}\right)  =\left(
0,10,20\right)  $\ \ (Right) $\left(  \theta_{1},\theta_{2},\theta_{3}\right)
=\left(  40,45,50\right)  $\ \ 
\end{center}

\bigskip

\bigskip

We estimate the LHS of the inequality:
\[
K(\theta_{1},\theta_{2},\theta_{3})=C_{12}+C_{23}-C_{13}.
\]
Table YY shows that the value of $K$ with respect to the $(\theta_{1}%
,\theta_{2},\theta_{3})=\left(  0,10,20\right)  $ and $(40,45,50)$. The value
of $K$ exceeding one is seen in several cases. 

\bigskip

\bigskip%

\begin{tabular}
[c]{cccccccccc}%
$\mathcal{X},\mathcal{Y},\mathcal{Z}$ & $\mathcal{C}_{A},\mathcal{C}_{A}$ &
$\mathcal{C}_{A},\mathcal{C}_{B}$ & $\mathcal{C}_{A},\mathcal{C}_{C}$ &
$\mathcal{C}_{B},\mathcal{C}_{A}$ & $\mathcal{C}_{B},\mathcal{C}_{B}$ &
$\mathcal{C}_{B},\mathcal{C}_{C}$ & $\mathcal{C}_{C},\mathcal{C}_{A}$ &
$\mathcal{C}_{C},\mathcal{C}_{B}$ & $\mathcal{C}_{C},\mathcal{C}_{C}$\\
$\mathcal{C}_{A}$ & $1.000$ & $1.214$ & $1.047$ & $0.870$ & $1.083$ & $0.917$
& $1.036$ & $1.250$ & $1.083$\\
$\mathcal{C}_{B}$ & $0.917$ & $1.130$ & $0.964$ & $0.786$ & $1.000$ & $0.833$
& $0.953$ & $1.167$ & $1.000$\\
$\mathcal{C}_{C}$ & $0.917$ & $1.130$ & $0.964$ & $0.786$ & $1.000$ & $0.833$
& $0.953$ & $1.167$ & $1.000$%
\end{tabular}

{\bf Table 1:} the triple of angles $\left(0,10,20\right).$  The values of $K$ for 
various combinations of contexts. For the contexts $\left(  \mathcal{C}_{A},\mathcal{C}%
_{C},\mathcal{C}_{B}\right),$  $K$ approaches its maximal value.

\bigskip%

\begin{tabular}
[c]{cccccccccc}%
$\mathcal{X},\mathcal{Y},\mathcal{Z}$ & $\mathcal{C}_{A},\mathcal{C}_{A}$ &
$\mathcal{C}_{A},\mathcal{C}_{B}$ & $\mathcal{C}_{A},\mathcal{C}_{C}$ &
$\mathcal{C}_{B},\mathcal{C}_{A}$ & $\mathcal{C}_{B},\mathcal{C}_{B}$ &
$\mathcal{C}_{B},\mathcal{C}_{C}$ & $\mathcal{C}_{C},\mathcal{C}_{A}$ &
$\mathcal{C}_{C},\mathcal{C}_{B}$ & $\mathcal{C}_{C},\mathcal{C}_{C}$\\
$\mathcal{C}_{A}$ & $0.055$ & $0.099$ & $0.015$ & $0.589$ & $0.633$ & $0.549$
& $0.339$ & $0.383$ & $0.299$\\
$\mathcal{C}_{B}$ & $0.339$ & $0.383$ & $0.299$ & $0.873$ & $0.917$ & $0.833$
& $0.623$ & $0.667$ & $0.583$\\
$\mathcal{C}_{C}$ & $0.589$ & $0.633$ & $0.549$ & $1.123$ & $1.167$ & $1.083$
& $0.873$ & $0.917$ & $0.833$%
\end{tabular}
{\bf Table 2:} the triple of angles $\left(  40,45,50\right).$  The values of $K$ for 
various combinations of contexts. For the contexts $\left(  \mathcal{C}_{C},\mathcal{C}%
_{B},\mathcal{C}_{B}\right),$  $K$ approaches its maximal value.

\bigskip

\section{Statistical analysis}

We start from the random variable:%
\[
K=Q_{1}Q_{2}+Q_{2}Q_{3}-Q_{1}Q_{3}%
\]
Here, $K$ takes $-4,-2,0$ or $+3$ since $Q_{i}$ takes $+1$ or $-1$. The
probability distribution of $K$ is not known, but it has mean value $\mu$ and
variance $\sigma^{2}$ and their statistical estimates can be found. To find the confidence interval in such a situation,
we apply the simplest method of nonparametric statistics, namely,
the method based on the {\it Chebyshev inequality}.    
(However, from the very beginning we remark that this method gives us only rough estimate for the confidence
interval.) This method was recently used  \cite{Zeilinger} for analysis of statistical data from the Vienna-test  
for the Bell-type inequality, the Eberhard inequality, which finally closed the fair sampling loophole. In this test, because of the presence
of slight drift depending on experimental setting, one cannot assume Gaussianity of data and it seems that usage of the Chebyshev 
inequality is the simplest way to resolve this problem.  

We can apply the Chebyshev
inequality to sample mean of $K$
\[
P\left(  \left\vert m-\mu\right\vert >c\right)  \leq\frac{\sigma^{2}}{nc^{2}}%
\]
with positive constant $c$. Here, $m$ is a sample mean of independent random
variables $K_{1},...,K_{n}$:
\[
m=\frac{K_{1}+K_{2}+...+K_{n}}{n}%
\]
Although we do not know the value of $\sigma^{2}$, we can estimate $\sigma
^{2}$ with unbiased sample variance. Then $\mu$ is estimated by $m$ with
confidence interval $\left[  m-c,m+c\right]  $. 

We take the $80\%$ confidence level.
In the case that the order of the contexts is $\left(  \mathcal{X}%
,\mathcal{Y},\mathcal{Z}\right)  =\left(  \mathcal{C}_{A},\mathcal{C}%
_{C},\mathcal{C}_{B}\right)  $, and the angle $\theta=0,10,20$ (This case is
maximum value of $K$) , we estimate the value of $K$ as follows.%
\[
K=1.250\pm 0.213
\]
Statistical analysis shows that the violation of the LG-inequality is statistically significant. 
However, it is clear that one has to perform better experiments to get a higher level of violation.
The main problem of the present experiment is that the sample is not large enough. It seems that 
this will be the problem of all Bell-type tests in cognitive science. All such inequalities are based
on correlations, so samples of students have to be sufficiently large: to calculate correlations (this needs human recourses)and get 
statistically significant violation (additional human recourse) of the corresponding inequality for such correlations.

\section{Concluding remarks}
\label{concl}

A violation of the contextual LG inequality by statistical data collected for observations of  the Schr\"oder stair rotated for different 
angles supports the {\it contextual cognition paradigm} presented in the series of works \cite{K1},\cite{OH}--\cite{OH2}, \cite{AS}. 
Our experimental statistical data is fundamentally contextual.\footnote{In fact, it violates even 
condition of marginal selectivity. We interpret violation of marginal selectivity as one of signs of contextuality. Of course, it is important 
to approach violation of the LG-inequality in combination with marginal selectivity. And this is a delicate issue. Up to authors' knowledge, 
even in physics the experimental situation is not completely clear. Practically all publications on violation of the Bell-type inequalities
represent only  the values of correlations (or even only the values of their linear combinations) and it is impossible   
to check condition of marginal selectivity. Only two authors, A. Aspect (in his PhD-thesis \cite{ASP}) and G. Weihs (open source, which 
was later removed),  presented count-data for which it is possible 
to check  marginal selectivity. And (surprisingly) this condition is violated \cite{ADNR}. Thus the question whether in physics combination of violation of 
the Bell-type inequalities and marginal selectivity was approached is open (mainly because experimental groups do not want to share 
count-data). We can speculate that in physics marginal selectivity would be never approached, at least in combination 
with violation of the Bell-type inequalities. Marginal distributions would always depend on experimental settings. However, the degree 
of such dependence can be considered as ``sufficiently small comparing with the degree of violation of the corresponding inequality.'' Of course, the latter
statement has to be presented in terms of statistical analysis.}  The brain 
does not have a priori prepared ``answers'' to the question about the R/L structure of   the Schr\"oder stair for the fixed angle $\theta.$ 
Answers are generated depending on the mental context. Thus mental realism is a kind of contextual realism, cf. \cite{KHR_CONT2}. 
There are practically 
no (at least not so many), so to say, ``absolute mental quantities'', ``answers'' to the same question vary essentially depending on context.
This conclusion is not surprising in the framework of cognitive science and psychology, where various framing effects are well known. 
Thus the main contribution of this paper is the demonstration of applicability of a statistical test of contextuality borrowed from quantum physics.
We also can consider this study as a step towards creation of unified mathematical picture of the world: physical and mental phenomena 
can be described by the same equations, cf. \cite{K1}. 

Such studies on usage of the standard quantum mechanical tests in other domains of science also contribute to foundations of quantum physics,
since they can be used as at least indirect arguments supporting some interpretations of outputs of these tests for physical systems.
In our case, the viewpoint that contextuality and not nonlocality is the basic source of nonclassical probabilistic behavior of 
cognitive systems, see also \cite{ADB}, may be used to support local contextual models of quantum physical phenomena. 

\section{Appendix: Kolmogorov probability model and Bell type inequalities}
 
\subsection{Kolmogorov axiomatics, probability space}
\label{KKK}

We start this section with rather long introduction to the classical measure-theoretic model of probability theory\cite{K}. Our aim
is to show that the problem of construction of common probability measure was one of the basic problems even in classical probability
theory, namely, invention of stochastic processes was based on such a special construction (Kolmogorov's theorem \cite{K}). Positive 
solution of this problem in the case of stochastic processes played an important role in the formation of the present ideology of classical 
probability: observables have to be represented by random variables on  common probability space. However, in the case of quantum 
observables this is not true.  

Classical probability theory is based on the model of
A. N. Kolmogorov \cite{K}.  Its basic notion is {\it probability space}, 
a triple ${\cal P}= (\Omega, {\cal F}, P),$ where $\Omega$ is a set, 
$P$ is a probability measure, and  ${\cal F}$ is a collection of subsets of $\Omega$ on which 
probability is defined.\footnote{This is a $\sigma$-algebra (``$\sigma$-field''): a system of sets which is closed with respect to countable 
unions, intersections, and the operation of complement.} In the Kolmogorovean model \cite{K} an {\it observable}, say $a,$
is represented by a {\it random variable}\footnote{We remark that the Kolmogorovean model is not 
simply a mathematical theory. In the same way as the Euclidean model
provided mathematical formalization of geometry of physical space, the Kolmogorovean model provided 
mathematical formalization of the theory of  measurements of random variables.
Euclid formalized such heuristic notions as point, straight line, plane, angle...; Kolmogorov formalized 
such notions as event, probability, random observable (=random variable).
In section \ref{geometry} we shall 
come back to comparison of the roles played by the 
Euclidean model of geometry of space and the Kolmogorovean model of random measurements in physics. 
}:  a map $a: \Omega\to {\bf R}$  such that, for each interval, 
its pre-image, $\{\omega \in \Omega: a(\omega) \in [\alpha, \beta) \},$ 
belongs to ${\cal F.}$ Its the probability distribution  is defined as 
$p_{a}(A) = P(\omega: a(\omega ) \in A).$

A system of observables is represented by a vector of random variables $a=(a_1,..., a_n).$ Its probability distribution is defined  
as 
\begin{equation}
\label{PD7}
p_{a}(A_1\times \cdots \times A_n) = P(\omega: a_1(\omega) \in A_1,\cdots, a_n(\omega) \in A_n).
\end{equation}
%We remark that, for any subset of indexes $i_1,...,i_k,$ the probability distribution of the vector 
%$(a_{i_1},...,a_{i_k}),$ $p_{a_{i_1},..., a_{i_k}}(A_{i_1}\times...\times A_{i_k}) = 
%P(\omega \in \Omega: a_{i_1}(\omega) \in A_{i_1}, ..., a_{i_k}(\omega)  \in A_{i_k}),$  can be obtained 
%from $p_a$ as its marginal distribution.

The notion of a random vector is generalized to the notion of a {\it stochastic process.} 
Suppose that the set of indexes is infinite; for example,  $a_t, t \in [0,+\infty).$ Suppose that, for 
each finite set $(t_1...t_k),$ the vector $(a_{t_1} ... a_{t_k})$ can be observed and its probability 
distribution $p_{t_1...t_k}$ is given. By selecting $\Omega_{t_1...t_k}= {\bf R}^k,$ 
$P_{t_1...t_k}= p_{t_1...t_k},$ and ${\cal F}$ as the Borel $\sigma$-algebra,
 we obtain the probability space ${\cal P}_{t_1...t_k}$
describing measurements at points $t_1...t_k.$   
At the beginning of 20th century
the main mathematical  question of probability theory was whether it is possible to find a single probability space 
${\cal P} =(\Omega, {\cal F}, P)$ such that all $a_t$ be represented as random variables on this space and all probability 
distributions $p_{t_1...t_k}$ are induced by the same $P:$ 
$$
p_{t_1...t_k}(A_1\times \cdots \times A_{k}) =P(\omega \in \Omega: a_{t_1}(\omega) \in A_{1},  , a_{t_n}(\omega)  
\in A_{n}).
$$ 
Kolmogorov found natural conditions for the system of measures $p_{t_1...t_k}$ which guarantee existence 
of such a probability space, see \cite{K}.\footnote{We just remark that 
the $\Omega$ is selected as the set of all trajectories $t \to \omega(t).$ 
The random variable $a_t$ is defined as $a_t(\omega) =\omega (t).$
Construction of the probability
measure $P$ serving for all finite random vectors is mathematically advanced and going back to construction of the Wiener 
measure on the space of continuous functions.}  
And during the next 80 years analysis of properties of (finite and infinite) families of random variables defined on one fixed  probability 
space was the main activity in probability theory.

\subsection{Kolmogorovean formalization of Bell's argument}

Although J. Bell did not formulate his argument in the terms of a probability space,  
the problem of local realism for quantum observables was formalized in complete accordance 
with  classical probability theory. In the Bell framework $\Omega$ is selected 
as the set of hidden variables $\Lambda;$ local realism is equivalent to mathematical presentation of observables
by random variables $\lambda \to a_\alpha(\lambda),b_\beta(\lambda).$  Here $a_\alpha, \alpha=\alpha_1, \alpha_2,$ 
are observables depending on the parameter $\alpha,$ experimental settings, at ``Alice's lab'' and
 $b_\beta, \beta= \beta_1, \beta_2,$ 
are observables depending on the parameter $\beta,$ experimental settings, at ``Bob's lab''.   
In the framework of the Kolmogorov probability model
the Bell inequality is a theorem. A. Fine rigorously proved \cite{F}
that the Bell inequality is satisfied iff the common 
probability space for random variables representing quantum observables does exist. 

In the LG inequality (which a special form of contextual Bell inequalities) there is a single 
observable $a_t$ depending on the time parameter. Kolmogorovness means that all these observables 
can be represented by random variables, $\lambda \to a_t(\lambda),$ on common Kolmogorov 
probability space.

\subsection{Models: (non-)Euclidean geometry and\\ (non-)Kolmogorovean probability}
\label{geometry}
Although the main stream in classical probability was, so to say, the ``common probability space stream'', 
we can point to a few attempts to swim against this stream, e.g., \cite{KHR_CONT1},  \cite{KHR_CONT}, \cite{KHR_CONT1a}, \cite{KHR_CONT2} 
 \cite{OHz}, \cite{Dzhafarov},  \cite{Accardi}--\cite{DZ2}. 

Now we come back to comparison of mathematical formalizations of geometry of physical space and random observations.
Since the work of Lobachevsky (and Gauss and Boyai), mathematicians understood that there exist different 
mathematical possibilities for representation of space geometry. We remark that already Lobachevsky and Gauss studied the 
problem of adequacy of the Euclidean model to physical reality. Lobachevsky proposed some astronomic tests, Euclidean contra Lobachevsky
geometries; in Germany Gauss (who had some administrative obligations to measurements of land) performed measurement of angles of 
a huge triangle formed by three mountains. The latter test confirmed that at least locally we live in Euclidean space. Later Riemann 
formulated the general principles of  geometry which played the fundamental role in mathematical representation of Einstein's
general relativity. (Lobachesky geometry was used in special relativity). 

We emphasize that mathematicians understood 
long before physicists (who were at that time completely busy with Newtonian physics based on the Euclidean geometry) that
the Euclidean geometry is one of possible models of space. On the basis of such an experience collected in the mathematical community
physicists were not astonished by the appearance of  Minkowski space in special relativity and then (pseudo-)Riemann space in general relativity.
Here the general approach of D. Hilbert on axiomatization of physics (also known as Hilbert's sixth problem) 
was respected. 

Development of mathematical formalization of probability was very different. Mathematicians (with a few 
exceptions) did not question the Kolmogorov axiomatics. The first non-Kolmogorovean model  
was elaborated in physics as a part of new physical theory -- quantum mechanics. And in classical probability community quantum probability 
is still not recognized  as a probability theory, but as some exercises in noncommutative algebra. Therefore in probability it is 
more difficult than it was in the case of geometry: any mathematical model, including the Kolmogorovean model, has a restricted domain of application.
Quantum phenomena simply showed that one special model of probability cannot be applied. From the viewpoint of Hilbert axiomatization of physics
this is the end of the story, i.e., one need not search additional ``explanations'' of non-Kolmogorovness, one simply has to find
a new appropriate mathematical model of probability and proceed with such a model. Thus from such a viewpoint the Bell argument is not about
locality and realism, but about inadequacy of the Kolmogorov model.\footnote{We point out that one of the problems slowing clarification of Bell's argument is that here typically probability 
(both classical and quantum) is not treated
in the axiomatic framework. There is a prejudice that ``probability is probability'' and it can be understood heurictically without going 
to mathematical axiomatization. However, the positive experience of physical applications of the axiomatic mathematical models of geometry
tells us that this is the most fruitful way even for  probabilistic applications.  (Nowadays in physics  nobody would work in the framework 
of  ``heuristically 
understandable geometry''.)} (Mathematical foundation of non-Kolmogorov probability theory was
discussed in the books \cite{KHR_CONT}, \cite{VOH}, \cite{NEW}.)
 
Let us again make a comparison with geometry. 
Did Einstein try to ``explain'' appearance of (pseudo-)Riemannian geometries in general relativity, 
instead of one special geometry (Euclidean)? Not at all, he simply
identified (pseudo-)Riemannian geometry with physical space.

Thus experimental tests of Bell's inequality can be considered as tests of adequacy of the traditional Kolmogorov model to 
quantum physical phenomena (cf. with aforementioned Gauss test of the Euclidean model).

Typically adherents of the non-Kolmogorovness viewpoint on violation of the Bell inequality consider physics
(in the spirit
 of the Hilbert program of axiomatization of physics) as a collection of mathematical models formalizing various natural 
phenomena. From the very beginning it is assumed that any such a model  has a restricted domain of applications. A physical experiment of 
which  the output cannot be described by a model under application is considered as a signal for creation of a new {\it mathematical} model, e.g.,
the Euclidean model could not be used for special relativity and new non-Euclidean models were explored (the Lobachevsky  in special relativity
and pseudo-Riemannian geometries in general relativity.)

\subsection{Reconstructing Kolmogorovness from contextuality}
\label{HHH}

The main message of this paper is that experimental statistical data collected 
in cognitive science and psychology are fundamentally contextual. In general for each experimental context data 
is described by its own Kolmogorov probability space.\footnote{We remark that this was the original viewpoint of Kolmogorov. In second section 
of his seminal monograph \cite{K} in which he formulated the axiomatics of classical probability theory he emphasized
the correspondence: experiment $\to$ probability space. Unfortunately, this message of the creator of probability theory 
was completely forgotten by his followers who completely ignored multi-space structure of experimental studies, see \cite{INT} for 
the detailed discussion. Of course, Kolmogorov by himself was excited by the possibility to unify probability spaces corresponding to 
measurements of a stochastic process for finite sequences of instances of time, see section \ref{KKK}. One cannot exclude 
that he was sure that such unification is possible for any kind of data. Although he had never pointed to such a 
possibility, some indirect signs supporting such a hypothesis about his views can be found; in particular, people from 
his close circle reacted very negatively to the attempt of Vorobj'ev \cite{V} to proceed with multi-space approach in a series 
of applications -- game theory, optimization theory. We remark that, in particular, Vorobj'ev \cite{V} derived (all possible) Bell type 
inequalities (for any number of variables yielding any fixed number of values). He used these inequalities as tests of Kolmogorovness 
of probabilistic data.}

\medskip

{\it Is it possible to unify these spaces in some way?} 

\medskip

One possibility is to use quantum probability and the  formalism of complex Hilbert space. 
This is so to say nonclassical unification. {\it  Surprisingly it is even possible to unify 
these probability spaces classically, i.e., to embed them into single ``big Kolmogorov 
space''!} Such classical unification is based on taking into account randomness of realizations
of contexts. In this ``big probability space''  the original probabilities appear as conditional probabilities with respect to various 
contexts. 

The first version of such unification was presented in the paper \cite{Astrid} (see
\cite{Astrid1} for better structured presentation), where the probabilistic data collected in experiments
to determine the EPR-Bohm correlations were embedded into single Kolmogorov probability space.
(This construction is evidently generalized to statistical data collected for any family of experimental 
contexts.)\footnote{See also recent studies \cite{DZQ}--\cite{DZQ2}.}

The reader may ask: Why do you emphasize non-Kolmogorovness in such a situation? 

Although ``big Kolmogorov space'' unifying data collected for different contexts exists,
it cannot be used so successfully as quantum probability. The latter describes all possible
experimental contexts homogeneously. In the classical approach based on reconstruction of Kolmogorovness
from contextuality one has to take into account randomness of realizations of concrete contexts. Of course,
one can construct huge Kolmogorov space unifying all possible contexts and all possible types of randomness 
for them. However, it would be practically impossible to work with such space.  
 
Another problem of the Kolmogorovian unification of contextual experimental data is that by taking into account randomness
of realizations of contexts (e.g., for the Bell-type experiments, how often each pair of orientations of polarization beam splitters 
is realized in the concrete experiment) we lost the internal description of data: experimenter's ``free will'' 
(to use this or that experimental context for the next trial)
also has to be taken into account.  This is a complex interpretation problem related to construction of ``big Kolmogorov spaces'',
see \cite{Astrid1} for discussion. Quantum mechanics provides description of experimental probabilities without taking into account 
randomness of realizations of experimental contexts; in this sense the quantum description can be treated as a kind of intrinsic 
description. And data can be ``intrinsically'' Kolmogorovian or non-Kolmogorovian (although, as we emphasized in 
this section, it is always possible to make these data Kolmogorovian ``externally'').

 \bigskip 

{\bf Acknowledgment:} We would like to thank H. Atmanspacher, I. Basieva,  N. Watanabe, and I. Yamato for discussions. 
This work was partially supported by MPNS COST Action MP1006 (Fundamental Problems in Quantum Physics) and visiting fellowships 
(A. Khrennikov) to  Center of Quantum BioInformatics, Tokyo University of Science and 
Institute for Quantum Optics and Quantum Information, Austrian Academy of Sciences.

\end{document}